\documentclass[usegraphicx]{mn2e}
\title[Star--disc interactions in a galactic centre]{Star--disc
 interactions in a galactic centre and oblateness of the inner stellar cluster}
\author[L.~\v{S}ubr, V.~Karas, and J.-M.~Hur\'e]{L.~\v{S}ubr,$\!^{1}$
 V.~Karas,$\!^{2,1}$ and
 J.-M.~Hur\'e$^{\,3,4}$\medskip\\
 $^1$~Astronomical Institute, Charles University,
  V~Hole\v{s}ovi\v{c}k\'ach~2, CZ-180\,00~Prague, Czech~Republic\\
 $^2$~Astronomical Institute, Academy of Sciences,
  Bo\v{c}n\'{\i}~II, CZ-141\,31~Prague, Czech~Republic\\
 $^3$~Observatoire~de~Paris-Meudon, Place~Jules~Janssen, F-92195~Meudon, 
 France\\
 $^4$~Universit\'e Paris~7 (Denis Diderot), 2~Place Jussieu, F-75251 Paris, 
 France
}
\date{Accepted 03 August 2004; Received 25 March 2004}
\pagerange{\pageref{firstpage}--\pageref{lastpage}}

\newcommand{\bs}{\!\!\!\!} 
\newcommand{\rd}{{\rm d}} 
\newcommand{\avg}[1]{{\langle #1 \rangle}}  
\newcommand{\myder}{{\mbox{d}}}
\newcommand{\md}{M_{\rm{}d}}

\newcommand{\mbh}{M_{\bullet}}
\newcommand{\sigmad}{\Sigma_{\rm{}d}}
\newcommand{\sigmas}{\Sigma_{\ast}}
\newcommand{\rg}{{R_{\rm{}g}}}
\newcommand{\rh}{{R_{\rm{}h}}}

\newcommand{\etal}{{\rm{}et~al.\ }}
\newcommand{\idxt}[1]{{\mbox{\tiny #1}}}
\newcommand{\idxs}[1]{{\mbox{\scriptsize #1}}}
\newcommand{\imf}{{{\cal{N}}_0}}
\newcommand{\mmf}{{\cal{N}}}
\newcommand{\betaan}{\beta_{\rm{a}}}
\newcommand{\Mdot}{\dot{M}_{\rm s}}
\newcommand{\Kdot}{\dot{K}_{\rm s}}
\newcommand{\tincl}{t_{\rm incl}}

\begin{document}
\label{firstpage}
\maketitle
\begin{abstract}
Structure of a quasi-stationary stellar cluster is modelled assuming
that it is embedded in the gravitational field of a super-massive black 
hole. Gradual orbital decay of stellar trajectories is caused by the
dissipative interaction with an accretion disc. Gravitational field of the disc
is constructed and its effect on the cluster structure is taken into
account as an axially symmetric perturbation. Attention is focused on a
circumnuclear region ($r\la10^4$ gravitational radii) where the effects
of the central black hole and the disc dominate over the influence of an
outer galaxy. It is shown how the stellar system becomes gradually
flattened towards the disc plane. For certain combinations of the model
parameters, a toroidal structure is formed by a fraction of stars.
Growing anisotropy of stellar velocities as well as their segregation
occur. The mass function of the inner cluster is modified and it
progressively departs from the asymptotic form assumed in the outer
cluster. A new stationary distribution can be characterized in terms of
velocity dispersion of the stellar sample in the central region of the
modified cluster.

\end{abstract}
\begin{keywords}
 accretion, accretion discs -- galaxies: nuclei -- galaxies: active
 -- stellar dynamics
\end{keywords}

\section{Introduction} 
We discuss the impact of gaseous environment on the long-term stellar
motion in central regions of active galactic nuclei (AGN). We adopt the
view of galactic nuclei as composite systems with stars surrounding a
super-massive black hole (with a typical mass of
$10^6M_{\sun}\la\mbh\la10^9M_{\sun}$) and forming a rather dense cluster
(Begelman \& Rees 1978; Norman \& Silk 1983). Central number density of
the cluster stands as one of the model parameters that can reach values
of the order of $n_{\ast}\sim10^6\div10^7\,\mbox{pc}^{-3}$. We consider
the stars of the cluster as satellites orbiting the central black hole.
We also take into account the influence that the diluted interstellar
environment exhibits on the motion of stars on time-scales exceeding
$10^4$ dynamical periods. This allows us to address an important
question whether or not the presence of such medium could increase the
concentration of stars in certain regions of the nucleus.

Although we consider the region of direct gravitational influence, where
the gravity of the central black hole dominates over non-gravitational
forces, one should not neglect the impact of perturbing effects on
individual stellar trajectories and the overall cluster structure. We
assume that it is the gas in an accretion disc which acts as the main
perturbation. The disc medium affects the motion of stars either
directly by collisions between stars and gas particles at the point of
transition through the disc slab (the case of inclined trajectories), or
vicariously, via coupling between the stars and the disc medium
(embedded trajectories). These processes will be discussed in
combination with the effects of disc gravity.

Our present contribution follows a previous paper (Vokrouhlick\'y \&
Karas 1998) where the long-term evolution of stellar orbits was
studied including the case of an axially symmetric perturbation imposed
on the gravitational field to mimic the presence of a disc. It was shown
that the orbits experience abrupt changes of eccentricity and
inclination even if the disc mass is a tiny fraction of the central
mass, $\md\la10^{-3}{\mbh}$. In that paper properties of individual
orbits were investigated. A simplified approach to the evolution of the
central cluster was adopted in Rauch (1999) and Karas, \v{S}ubr
\& \v{S}lechta (2002). Stellar encounters were explored in this context
by Vilkoviskij \& Czerny (2002), using a semi-analytical approach to the
dense and flattened cluster. Here we add an astrophysically more 
realistic description of the self-gravitating disc, especially of
its gravitational field that superimposes with the central field, and
we study dynamics of the cluster in terms of its overall characteristics
(the shape and velocity dispersion).

The main new ingredients of the present investigation are threefold.
Firstly, a more elaborate description of the disc accounts for 
different processes and corresponding time-scales that dominate the 
system evolution at different distances from the core. Orbit behaviour
is governed mainly by the overall radial profile of the disc (especially
its surface density and vertical thickness) and by stellar
characteristics (mass and size). The interplay between these principal
parameters determines the hydrodynamical mechanism driving the
evolution --
collisions with the disc, density waves, or formation of gaps in the
disc.

Secondly, we introduce the distribution of stellar masses in the
cluster, $\mmf(M_{\ast})$. The cluster is supplied with a Salpeter type
initial mass function, but the resulting mass function is eventually
different because of selective orbital decay. For our purposes, stars
are characterized phenomenologically by their effective column density
$\sigmas{\equiv}M_{\ast}/\left({\pi}R_{\ast}\right)^2$ as a key paramater
that rules this effect.

Thirdly, we consider the distribution of stars in phase space. 
Simulations start with a spherically symmetric and stationary cluster
whose dynamics is dominated by the central black hole mass (Bahcall \&
Wolf 1976, {\sf{}BW} hereafter). This solution describes a
gravitationally relaxed  system of stars near a black hole with
phase-space probability density $f_0(a)\,\propto\,a^{1/4}$
($f_0(a)\,{\myder}a$ is the number of stars in a shell with semi-major
axis $a$ and width ${\myder}a$). We assume that the modified
distribution approaches $f_0(a)$ far away from the core, but it can be
modified in the inner parts of the system where also the importance of
the disc increases. Other self-consistent models have been discussed for
a cluster with a central black hole (e.g.\ Young 1980; Quinlan,
Hernquist \& Sigurdsson 1995; Takahashi \& Lee 2000; Alexander 2003) for
which the radial density profile turns out to be different but also a
power-law type at large radii. Furthermore, several authors (e.g.\ 
Rauch 1995; Freitag \& Benz 2002; Alexander \& Hopman 2003) have
developed numerical approaches taking stellar collisions into account in
a dense cluster near a black hole.

The paper is organized as follows. In the next section we describe the
ingredients of the model and time-scales involved. In
Section~\ref{results}, we show a modified quasi-stationary stellar
distribution that emerges in the inner regions of the cluster. We
discuss transition regions where the stars are gathered and we also
examine predicted velocity dispersions, mean spectral-line  profiles and
the modified mass function. Some of these properties may  have direct
observational consequences. Finally, limitations of the model are
discussed and conclusions are presented in Section~\ref{conclusions}. 

\section{The model}

\subsection{Components of the model}
Our exploratory model is based on a presumably obvious assumption of
dissipative processes prevailing in the center, where the disc surface
density is high, while gravitational relaxation is supposed to be in
control on the outskirts of the system behind the outer edge of the
disc. The adopted approach contains simplifications that help us to
proceed in the situation when $N$-body or equivalent methods are
still a challenge, especially if the dissipative environment, different
types of stars and non-spherical geometry of the system are to be taken
into account. The model consists of three components:
\begin{description}
\item[(a)]~{\em{}A super-massive black hole.} Its influence is
approximated by Newtonian gravitational field of a point mass $\mbh$.
Relativistic effects are introduced only by adopting
$\rg{\equiv}G{\mbh}/c^2\sim1.5\times10^{13}
\,M_8\,\rm{cm}$ ($M_8{\equiv}{\mbh}/10^8M_{\sun}$) as the smallest natural 
length-scale that appears in our considerations. The corresponding
dynamical time-scale is much shorter than all other time-scales in the
model and it does not stand explicitely in equations of the long-term
evolution hereafter.     
\item[(b)]~{\em{}An accretion disc.} In order that computations be 
well defined and easier to compare, we consider the standard
Shakura--Sunyaev disc (denoted as ``{\sf{}SS}''; e.g.\ Frank, King \& 
Raine 2002) and a model of weakly self-gravitating disc (Goldreich \&
Lynden-Bell 1965; Paczy\'nski 1978; Shlosman \& Begelman 1989). We 
assume that approximation of vertically averaged quantities is
sufficient for our analysis and we characterize the disc by surface
density $\sigmad(r)$ and geometrical semi-thickness $H(r)$ (functions of
radius in the disc mid-plane $z=0$). Other variables are
also relevant for star--disc interactions, e.g.\ radial velocity of mass
transport in the disc, temperature of the medium and speed of sound.
Gravitational field is determined numerically. With the disc we
associate time-scale $\tincl$, i.e.\ the interval on which stellar
orbits are inclined into the plane by the dissipative interaction.
\item[(c)]~{\em{}A nuclear cluster.} It is characterized by the 
distribution function $f_{\ast}$ that conforms initially to a
spherically symmetric and stationary state in the central potential
well. In our computational scheme, the cluster is divided in two
distinct subsets that are treated separately. In the {\em inner
cluster}, the stellar system becomes oblate and anisotropic due to the
dissipative interaction with the disc. Efficiency of this interaction
depends on the parameters of individual stars and, therefore, it is
sensitive to their masses, orbital inclinations, as well as to
properties of the dissipative environment. Gravitational relaxation is
neglected in the inner cluster, assuming that $t_\idxs{r}\gg\tincl$. The
{\em outer cluster}, on the other hand, does not evolve in our
computations and its structure is determined by gravitational relaxation
as a dominating process. The outer cluster acts as a reservoir from
which fresh stars are injected inwards, maintaining the overall steady
state.  
\end{description}

\begin{figure*}
\includegraphics[width=\textwidth]{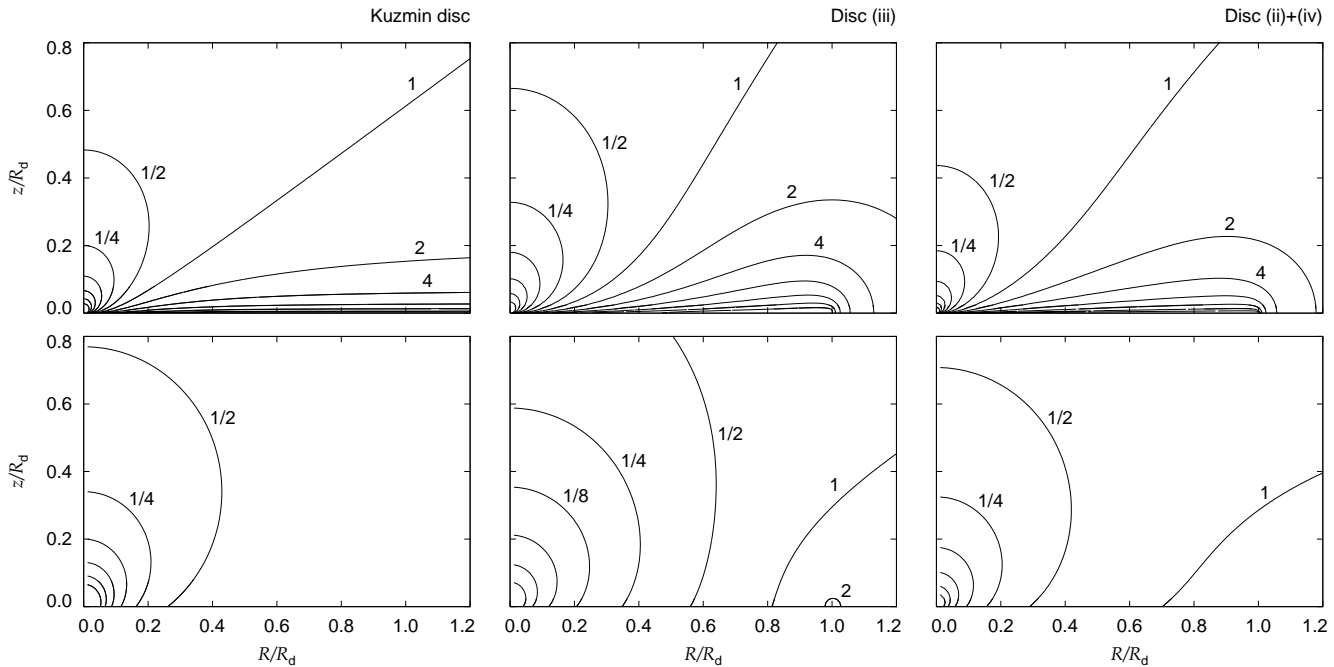}
\caption{Gravitational field of different models of a disc is shown in
comparison with the central field of the black hole. Ratio of the 
corresponding field components was computed and the resulting contours
plotted (in azimuthal section). The ratio of vertical ($z$) components
is shown in the top row (i.e.\
$F_{z,\rm{}disc}/F_{z,\rm{}centre}={\rm{}const}$), while the ratio of
radial (cylindrical $R$) components is in bottom
($F_{R,\rm{}disc}/F_{R,\rm{}centre}={\rm{}const}$). Obviously, near the
equatorial plane the vertical component of the disc gravity dominates.
Three representative types of discs have been selected for illustration,
as indicated above the frames. Contours are drawn in logarithmic spacing
and the values refer to $\mu=1$.}
\label{fig:kuzmin}
\end{figure*}

\subsection{An accretion disc with effects of self-gravity} 
\label{disc}  

We adopt three flavours of the standard model: (i)~radiation pressure
dominated standard ({\sf{}SS}) disc; (ii)~gas-pressure dominated
{\sf{}SS} disc with electron scattering; and (iii)~the same as the
previous case but with free-free opacity (Frank et al.\ 2002). Radial
dependence of the disc variables is expressed by power-law in all those
cases, in particular, the surface density can be written in the form
$\sigmad/\sigmas\,\propto\,\left(r/\rg\right)^s$. The values of index
$s$ are given in Table~\ref{tab:s_index} for different cases (see also
Karas \& \v{S}ubr 2001) and we will introduce also non-powerlaw
profiles later in this section. 

\begin{table}
\caption{Values of the power-law index $s$ characterizing the radial 
dependency of the disc surface density $\sigmad(r)$.
\label{tab:s_index}}
\begin{center}
\begin{tabular}{lccccc}
Model & (i) & (ii) & (iii) & (iv) & (v) \\ \hline
$s$ parameter & $3/2$ & $-3/5$ & $-3/4$ & $-15/7$ & $-5/3$
\end{tabular}
\end{center}
\end{table}

The standard scheme was supplemented by considering self-gravity 
according to Collin \& Hur\'e (1999, 2001) and Hur\'e et al.\ (1994, 2000).
Based on Toomre's criterion ($Q\sim1$) for onset of self-gravity,
AGN accretion discs are affected beyond critical radius
$\sim10^3\rg$ (i.e.\ $\sim0.01$~pc, depending on
model parameters). Farther out, at $\ga1$~pc from the  center,
the disc mass can reach a sizeable fraction of the central mass. For
example, the mass ratio of the disc (ii) with respect to the central
mass is $\mu\equiv\md/\mbh=0.3M_8^{6/5}\dot{M}_{0.1}^{3/5}\alpha_{0.1}^{-4/5}
\left(R_\idxs{d}/10^5\rg\right)^{7/5}$ where we $R_\idxs{d}$ is the
disc outer radius, $\dot{M}_{0.1}=\dot{M}/0.1\dot{M}_{\rm{}Edd}$,
$\alpha_{0.1}=\alpha/0.1$. Similarly, for case (iii) we have 
$\mu=0.1M_8^{6/5}\dot{M}_{0.1}^{7/10}\alpha_{0.1}^{-4/5}
\left(R_\idxs{d}/10^5\rg\right)^{5/4}$, suggesting  that the disc
attraction is strong enough to influence the motion of  stars
independently of the details of a particular model. Self-gravity
manifests itself by gathering material to the mid-plane, thereby
narrowing vertical thickness.

We considered the case of marginally stable self-gravitating discs. Two
examples, denoted as (iv) and (v) hereafter, are characterized by
$Q\sim1$ and they differ by assumption about elemental abundances: solar
metalicity is adopted in the case (iv) while it is set to zero in the
case (v). Because of differences in opacities, different density
profiles and values of the total disc mass are obtained. Typically, in
the outer region of a marginally stable disc, density decreases faster
compared to the standard case. For a detailed discussion of the adopted
models, see Hur\'e (2000) and references cited therein.

Finally, we considered composite discs, by which we mean a combination
of simple models. For example, case (ii)+(iv) is constructed by
concatenating two corresponding basic models. Notation indicates that
the standard {\sf{}SS} disc (ii) is assumed at small radii and matched
to a marginally self-gravitating disc (iv) at large distance from the
centre (density and other variables are continuous at the transition
radius). This approach reflects an obvious fact that simple models can
be valid only for a limited range of radius and provides setup for the
model that is rich enough to capture different possibilities. 

Figure~\ref{fig:kuzmin} illustrates three examples of the structure of
the gravitational field. To this aim we plot the ratio of field
components that correspond to the disc and to the central black hole
(contribution of the cluster stars is not included in this figure).
Different frames reveal the difference between the case of Kuzmin's disc
(employed in Vokrouhlick\'y \& Karas 1998) and two typical examples of
our discs, computed numerically for this paper. Even if the
gravitational field of the disc is perturbative ($\mu\ll1$), it turns
out that the vertical component of the disc field may successfully
compete with that of the centre in certain regions near the disc plane.
This is essential not only for the disc internal structure but we can
also expect stellar paths to be visibly dragged (on the long term),
especially those which follow a low-inclination orbit, i.e.\ almost
embedded in the disc.

The gravitational potential and the field due to disc are  computed
according to Poisson's integrals in which integrations are carried over
volume occupied by the gas, assuming axial symmetry.  We used the
splitting method for accurate evaluation (Hur\'e 2000). Furthermore, in
order to reduce duration of runs, we implemented a dynamical grid  in
the Poisson solver and we pre-generated the field components in a
Cartesian grid covering a rectangular region of 1~pc\,$\times$\,1~pc. In
some regions, close to the disc, the field lines are very distorted. We
found it convenient to define additional sub-grids which provide fine
resolution where needed. Field values between mesh points were obtained
by bilinear interpolation. Hereafter, we will illustrate results for
discs (iii) and (ii)+(iv), i.e.\ the same as in Fig.~\ref{fig:kuzmin}.
These two cases appear sufficient in order to demonstrate various stages
of the modified cluster. The models share the same black hole mass,
$\mbh=10^8M_{\sun}$, as well as other common parameters of the disc:
$R_\idxs{d}=10^4\rg$, $\dot{M}=0.1\dot{M}_\idxt{E}$, and $\alpha=0.1$.
Corresponding mass ratio is $\mu=1.2\times10^{-2}$ and
$\mu=2.9\times10^{-3}$ for discs (iii) and (ii)+(iv) respectively.

\begin{figure*}
\includegraphics[width=\textwidth]{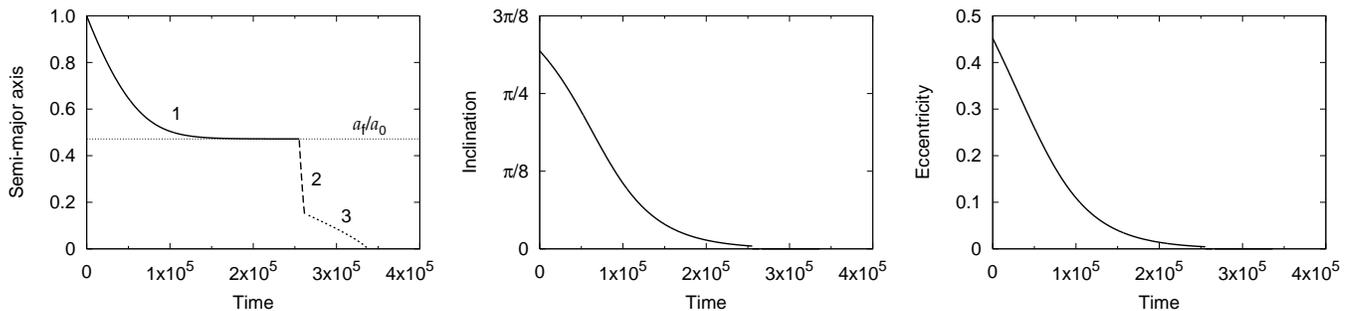}
\caption{Long-term evolution of orbital elements for a single star. 
Different modes of star--disc interaction are distinguished by line type
in the graphs, corresponding to the three subsequent phases of the mean
motion (details in the text): in phase~1 the regime of direct collisions
takes place (solid line); in phase~2 the regime of density waves occurs, 
in which $a$ drops more rapidly (dashed line); finally, phase~3
describes migration of a star tidally coupled to the disc (dotted
line). Time is scaled with respect to the initial dynamical period,
$T_0=2\pi\sqrt{G\mbh}\,a_0^{3/2}$; semi-major axis is expressed in terms
of the initial $a_0$.}
\label{fig:in_disc}
\end{figure*}

\subsection{Star--disc interactions}
Stars passing through the disc may create gaps and holes in some
regions, but we assume that the disc is not removed completely, implying
that the accretion rate cannot be too low and the cluster excessively
dense. We adopt the approximation of impulsive (i.e.\ instantaneous)
interaction between crashing stars and the disc (Syer, Clarke \& Rees
1991; Karas \& Vokrouhlick\'y 1994). In this way, stellar trajectories
slowly evolve and several phases can be distinguished in the course of
orbital evolution.

During the initial phase, as long as the stars are crashing on the disc
with non-zero inclination, the following combinations of orbital
elements $a$, $x\equiv\cos{i}$ and $y\equiv(1-e^2)$ remain constant:
\begin{equation}
C_1 = a \, y\, (1+x)^2 \quad\mbox{and}\quad
C_2 = \frac{(1-y)(1+x)^3}{1-x}
\label{eq:circular_orbit}
\end{equation}
(see \v{S}ubr \& Karas 1999). These two conserved quantities can be 
employed to determine the orbital evolution along with
a simple estimate of $\tincl$,
\begin{equation} 
\tincl = t_0
M_8\,\frac{\sigmas}{\Sigma_{\sun}} 
\left(\frac{a_0}{\rg}\right)^{3/2-s}\, {\rm yr}\,. 
\label{eq:t_x_circ1}
\end{equation}
Here, various model-dependent factors were absorbed in the term
$t_0{\equiv}t_0(a_0,e_0,i_0)$ that varies only slightly with $a_0$. For
example, in cases (ii),\,\ldots,\,(v), and for $e_0=0$, $i_0=\pi/2$, we
find $t_0=0.92$, $0.22$, $1.89\times10^{-6}$ and $9.40\times10^{-4}$,
respectively (these numbers were determined numerically).\footnote{It 
can be shown that $C_1$ and $C_2$ are 
strictly conserved for orbits with pericenter argument $\omega=\pi/2$,
but their variation for arbitrary $\omega$ is only weak.
In the case (i), $t_0$ would come out very small compared to other cases
because $\sigmad$ increases with $R$. However, this is the radiation
pressure dominated disc, which can be relevant only in the innermost
regions, and so it should not be used to describe the whole process of
inclining the trajectory into the disc. We avoid such behaviour by using
composite models of the disc as described above.} Because there is
considerable uncertainty in the actual value of the disc surface density
and since the cluster consists of various stellar types, ranging from
compact stars to giants, it is easy to realize that $\tincl$ must span
an enormous range over seven orders of magnitude in
eq.~(\ref{eq:t_x_circ1}).

Inclination and eccentricity decay exponentially at late stages when $i
\propto \exp[-t/\tau]$, $e\approx\sqrt{C_2}/(4i)$, while the semi-major
axis slows down and eventually reaches its asymptotical value
$a_\idxs{f}=C_1/4$. When inclination drops to a critical value (given by
$\tan{i}{\sim}H(r)/r$), the regime of direct collisions goes over to the
regime of stars entirely embedded in the disc. Since this moment, we
further distinguish two modes of inward migration through the
disc medium. If the stellar mass is large enough to create a gap (e.g.,
Lin \& Papaloizou 1986), the star becomes tidally coupled with the disc
and it sinks to the centre at roughly the same rate as the material of
the disc. Otherwise, excitation of density waves dominates and causes
radial migration to be faster, typically by order of magnitude (Ward
1986; Artymowicz 1994).

In Figure~\ref{fig:in_disc} we show a particular case of an orbit
evolving with the disc (iii) and $M_\ast=3M_\odot$, ensuring that the
star successively enters {\it all three modes of migration}. Starting
from $a_0=10^3\rg$, $e_0=0.45$ and $x_0=0.54$, the semi-major axis
converges to $a_\idxs{f}=0.47a_0$ along with the simultaneous decay of
eccentricity and inclination. The initial phase of star--disc collisions
is followed by the decay due to density waves. Finally, as the disc
parameters change with radius, the star eventually clears a gap in the
disc and the dominant mode of its radial migration is switched once
more.

\begin{figure*}
\includegraphics[width=\textwidth]{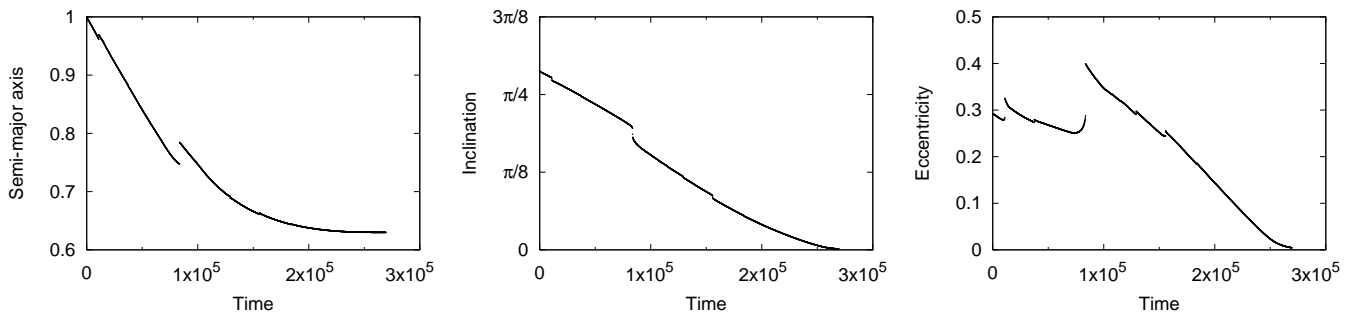}
\caption{Similar to the previous figure, but for stars moving in 
superposed gravitational field of the central black hole mass and an
axially symmetric thin disc. The spherical symmetry is now broken,
bringing new features into the orbit evolution. Namely, sudden steps in
the osculating elements occur at resonances between radial and vertical
oscillation frequencies. Again, semi-major axis is scaled with respect
to the initial value, $a_0$, and time is given in units of $T_0$ (only
a part of the whole evolution is shown for better clarity). Short-term
variations of the elements were filtered out in this graph.}
\label{fig:in_disc2}
\end{figure*}

Figure \ref{fig:in_disc2} illustrates an interesting effect that emerges
once the gravitational field of the disc is taken into account. The disc
introduces a non-spherical perturbation to the background gravitational
field of the central mass. As a consequence, a Kozai-type mechanism
causes occasional jumps of orbital parameters. A specific
form of the disc gravitational field is not particularly crucial for
the onset of these hops, but the dissipative orbital decay and
appropriate arrangement of time-scales are important (see Vokrouhlick\'y
\& Karas 1998 for more discussion). In fact, any non-spherical
perturbation, for example general-relativity effects of frame-dragging
near a rotating black hole, can potentially be responsible for similar
kind of oscillations. 

In the following sections we discuss integral properties of the whole
cluster, and therefore we will sum over individual stars. One should
bear in mind that orbital resonances can set stars on very eccentric
trajectory, increasing the capture rate. Hence the presence of the disc
may influence the overall cluster structure in its inner regions.

\subsection{The cluster} 
\label{sec:cluster}
Stationary state is maintained by a continuous inflow of fresh stars
from the outer reservoir and this is accomplished by distinguishing
between two populations of orbits: those which belong to the relaxed
cluster with a prescribed distribution $f_0$ and those which are
influenced by the disc and belong to the inner cluster. The latter
contribute by fraction $f_0\,\tincl/t_\idxs{r}$. The cluster as a whole
is superposition of these two populations; the corresponding
distribution will be denoted $f_\ast$ hereafter. Because the influence
of the disc on stellar orbits is greater at small distances and likewise
the effect of circularization increases inwards, therefore the
distinction between the two populations reflects also their spatial
dimensions.

First we introduce the distribution $f_0(a_0,e_0,x_0,\omega_0)$, which
characterizes the initial form of the cluster so far unperturbed by the
disc. It is determined by relaxation processes, acting on time-scale
\begin{equation}
t_\idxs{r}=\frac{\sigma^3}{G^2C\ln\Lambda\,M_{\ast}^2n_{\ast}},
\label{eq:t_relax_bound}
\end{equation}
where $\sigma$ is the velocity dispersion, $n_{\ast}$ the number density of
the stellar system, $\ln\Lambda$ is Coulomb logarithm and $C$ is a
constant (in usual notation, $C\ln\Lambda\sim10^2$; e.g.\ Spitzer 1987).

If relaxation was neglected in comparison with the dragging of stars by the
disc, one could write probability of finding a star on an orbit with
semi-major axis $a$ in the form 
\begin{equation}
 f_{\ast}(a;a_0,e_0,x_0,\omega_0)\,{\rd}a=
 {\rd}\Omega\,f_0(a_0,e_0,x_0,\omega_0)
 \,t_\idxs{r}^{-1} J^{-1}\,;
 \label{eq:part_na}
\end{equation}
where $J\equiv\|\rd{}a(a;a_0,e_0,x_0,\omega_0)/\rd{}t\|$ and 
$\rd\Omega$ is the phase space element. Meaning of this relation can be
clarified by integrating eq.~(\ref{eq:part_na}) over $a$ and obtaining
the total number of stars in the inner cluster which have been injected
from $\rd\Omega$. This number is equal to the fraction of stars in
$\rd\Omega$ multiplied by the ratio of drag period to the relaxation
period. At the same time, integration over the phase space of initial
parameters gives the desired probability density in the inner cluster:
\begin{equation}
 f_\idxs{in}(a)\,\rd a = \int_\Omega \rd \Omega_0 \,
 f_0(a_0, e_0, x_0, \omega_0) \,t_\idxs{r}^{-1}J^{-1}\,.
 \label{eq:supplied_density}
\end{equation}
The way in which the boundary zone is introduced between the inner and
the outer populations is the main advantage (and limitation) of the
adopted approach. It is assumed that $\Omega$ is vacated solely by
dissipative effects (interaction with the disc), while it is filled, at
a constant rate, solely via relaxation. In the numerical scheme this is
achieved by requiring $\tincl$ to be shorter  than relaxation time
(\ref{eq:t_relax_bound}) in the outer cluster, i.e.
\begin{equation}
\tincl \sim \frac{a_0}{|\dot{a}|} \leq t_\idxs{r},
\label{eq:drag_vs_relax}
\end{equation}
where $\dot{a}$ is the rate of orbital decay (we substituted the value
of $\dot{a}$ at initial time). Another condition is imposed by requiring
that the orbit intersects the disc,
\begin{equation}
R_\idxs{d} > R_0\equiv\frac{a_0\,(1 - e_0^2)}{1+e_0\,\cos\omega_0}.
\label{eq:geometry_condition}
\end{equation}

A Monte Carlo code was used to evaluate the integral on the
right-hand-side of eq.~(\ref{eq:supplied_density}). We followed 
$N\sim10^7$ orbits with randomly generated starting values of osculating
elements. Each time an orbit reaches a pre-determined value of $a_i$
(these are distributed in a logarithmical grid) the corresponding bin of
the probability-density array is increased by factor $|\dot{a}|^{-1}$.
This contributes by
\begin{equation}
f_i = \frac{1}{t_\idxs{r}} \frac{N_0}{N} \sum_1^{N_\idxt{c}} \left|
\frac{\rd a}{\rd t}(a;{\mathbf{}r},{\mathbf{}v}) \right|^{-1}\,
\end{equation}
to the distribution. Here, $N_0$ is the total number of stars within the
specified range of radii in the outer cluster, and $N_\idxs{c}$ is the
number of generated orbits intersecting the disc and satisfying
conditions (\ref{eq:drag_vs_relax}) and (\ref{eq:geometry_condition}).
In typical calculation, $N_\idxs{c}$ reaches a few percent of $N$.

The code enables us to further distinguish two subsamples of the inner
cluster. Stars on inclined orbits intersecting periodically the disc are
counted to a {\em{}dragged cluster}. Once a star is embedded entirely
into the plane of the disc and the mode of its radial migration is 
changed accordingly, we begin counting the star as a member of an
{\em{}embedded cluster}. It turns out that these two subsamples populate
different regions of the parameter space of the inner cluster, and so
the distinction appears plausible.

We have employed the {\sf{}BW} distribution to populate the outer
reservoir. Maxwellian velocity distribution is used as a boundary condition
specifying the dispersion $\sigma_\idxs{c}$ at zero binding energy surface. 
As mentioned above, $f_0(a_0)\,\propto\,a_0^{1/4}$ with $a_0$ being within
${\langle}a_\idxs{min},a_\idxs{max}{\rangle}$. We set 
$a_\idxs{min}=R_\idxs{d}$ and $a_\idxs{max}=R_\idxs{h}$, where
$R_\idxs{h}$ is radius of the black hole gravitational dominance,
\begin{equation}
\rh=G{\mbh}/\sigma_{\rm{}c}^2\sim 4.5M_8
\sigma_{200}^{-2}\,{\rm pc},
\label{eq:rh}
\end{equation}
written here in terms of $\sigma_{200}\equiv\sigma_{\rm{}c}/
(200\,{\rm{}km\,s}^{-1})$. Remaining orbital parameters, $x_0$, $e_0$
and $\omega_0$, are distributed randomly with equal probability, ensuring
spherical symmetry of the outer cluster. Corresponding spatial number density is
\begin{equation}
n_{\ast}(r)\sim(r/R_\idxs{h})^{-7/4}\,n_0\,.
\label{eq:bw_nr}
\end{equation}
The velocity dispersion of the {\sf{}BW} cluster scales roughly as
$\sigma^2{\sim}G\mbh/r$. 
With help of eqs.\ (\ref{eq:rh}) and (\ref{eq:bw_nr}), relaxation
time (\ref{eq:t_relax_bound}) can be rewritten as
\begin{equation}
 t_\idxs{r}= 10^{8}n_6\, M_8^{7/8} \biggl( \frac{M_\ast}{M_\odot}\biggr)^{-2}
 \biggr(\frac{r}{\rg} \biggr)^{\frac{1}{4}} {\rm yr}
\label{eq:t_relax_bw}
\end{equation}
with $n_6\equiv(n_0/10^6)\,$pc$^{-3}$. 

Next, we need to specify a relation between $\sigma_{\rm c}$ and $\mbh$ that
would hold outside $R_\idxs{h}$. We therefore adopt the empirical relation 
(Ferrarese \& Merritt 2000; Gebhardt et al.\ 2000),
$\log_{10} M_8=\alpha+\beta \log_{10}\sigma_{200}$
with $\alpha=0.13$ and $\beta=4$ (cf.\ Tremaine et al.\ 2002). In our model,
the log-log formula for $\sigma_{\rm c}(\mbh)$ is used to define the distribution
of stars in the bulge overlaying our nuclear cluster and this way
it unavoidably influences parameters derived from any observation. 
Otherwise, the forms of this relation in the two regions are quite independent
of each other. Notice that simplicity is one of the
reasons for adopting the {\sf{}BW} outer cluster, but it is not
necessary for the method as such. Likewise, different distributions of
eccentricities can be accommodated. For example, we tested
$f_{\ast}(e)\,\propto\,e$ and $f_{\ast}(e)\,\propto\,\exp(e)$ without
much impact on the results, i.e.\ on the final distribution of the
system. 

The spatial size of the inner cluster is determined by the interplay between
gravitational relaxation and dissipative effects in the disc. To
estimate the size, two length-scales are naturally involved.  Firstly,
from eq.~(\ref{eq:geometry_condition}) it is the disc radius
$R_\idxs{d}$ which limits the region of star--disc collisions. Secondly,
we notice that $t_\idxs{r}(r)$ increases with radius at slower rate than
$\tincl(a)$. Hence, by substituting $a{\rightarrow}r$  and setting
$t_\idxs{r}=\tincl$ we define a characteristic radius
\begin{equation}
 \left(\frac{R_\idxs{i}}{R_\idxs{i$_0$}}\right)^{\frac{5}{4}-s}
 \!\!\!=n_6^{-1}M_8^{-1} 
 \frac{\sigmas}{\Sigma_{\sun}}
 \left( \frac{M_\ast}{M_{\sun}} \right)^{-2} 
 \left( \frac{R_\idxs{h}}{10^6\rg}
 \right)^{-\frac{7}{4}} .
 \label{eq:r_inner}
\end{equation}
For our exemplary disc models, one finds the following
values:\footnote{In the case (i) the estimate of drag time comes out
more complicated, but we can safely consider that $t_\idxs{r}\gg\tincl$
for any reasonable choice of model (i) parameters.}
(ii)~$R_\idxs{i$_0$}=4.73\times10^4\rg$, (iii)~$4.36\times10^4\rg$,
(iv)~$1.68\times 10^4\rg$, and (v)~$9.80\times 10^3\rg$.
Eq.~(\ref{eq:r_inner}) is useful for analytical estimates of the cluster
size. The smaller of the two radii $R_\idxs{i}$ and $R_\idxs{d}$
corresponds to the size of the inner cluster. This estimate is in agreement
with numerical simulations. 

Finally, we modified the above-described scheme and tested a model with
the initial distribution $f_0$ being set to be Maxwellian in velocities.
By comparing the results based on both approaches we found qualitative
agreement between these two different formulations (see Karas et al.\
2002 for details). 

\section{Results}
\label{results}
\subsection{Radial structure of the modified cluster}
\label{results1}
\begin{figure}
\includegraphics[width=\columnwidth]{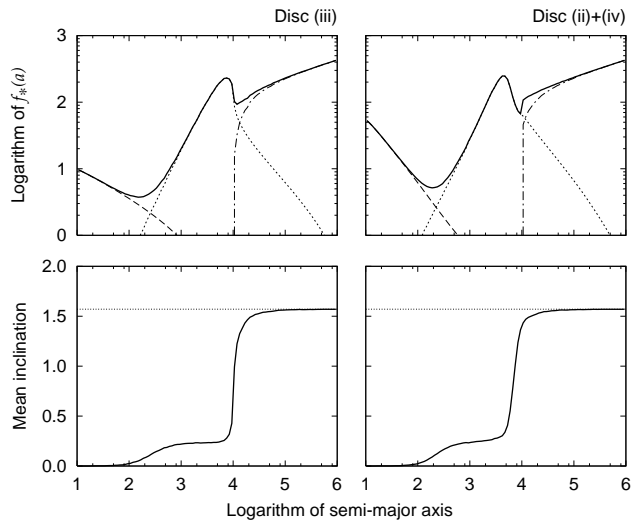}
\caption{Typical examples of the structure of a cluster resulting from
its interaction with different models of an accretion disc. Top panels:
distribution $f_{\ast}(a)$ of stars (solid line, in arbitrary 
normalisation) residing in a shell of given semi-axis ($a$ in units of $\rg$).
Dashed, dotted, and dash-dotted lines stand for the three sub-components
of the cluster, i.e.\ the embedded, dragged, and relaxed (outer)
population. Bottom panels: mean
orbital inclination $\avg{i}$ throughout the whole cluster, exhibiting
the effect of flattening below $10^4\rg$.
\label{fig:radial_profile}}
\end{figure}

Let us first discuss the impact of star--disc interactions on the 
radial distribution of stars. Figure~\ref{fig:radial_profile} provides
connection between the properties of individual orbits and the overall
evolution of the cluster. Parameters of the disc have been chosen
identical as in Fig.~\ref{fig:in_disc}, in particular, the outer edge
$R_\idxs{d}=10^4\rg$. Additional parameter is the stellar density, 
$n_0\sim10^6{\rm{}pc}^{-3}$. In this case, radius $R_\idxs{i}$, defined
by eq.~(\ref{eq:r_inner}), is approximately $2\times10^4\rg$, i.e.\
twice the disc outer radius. Therefore, the transition from the inner to
the outer cluster occurs at $r=R_\idxs{d}$. Clearly, the transition area
introduced in this way is not a sharp value of radius -- it depends
especially on eccentricity distribution of the stellar system,
$R_\idxs{d}$ of the disc and $\tincl/t_\idxs{r}$ of the both components.
The resultant number density, $f_{\ast}(a)$, is closely matched by a
broken power-law with a slope depending on the dominant regime of
star--disc interaction. Notice that the asymptotic form is
$f_{\ast}\propto\,a^{1/4}$ for large $a$.

\begin{figure}
\includegraphics[width=\columnwidth]{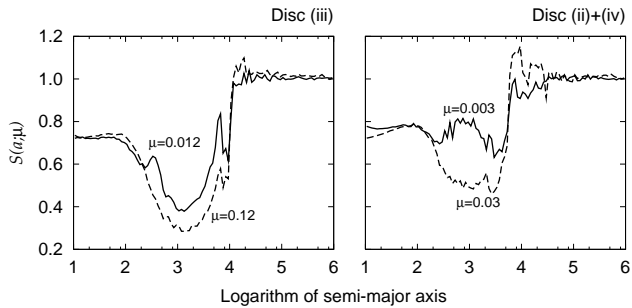}
\caption{Ratio $S(a;\mu)$ of two $f_\ast$ distributions obtained 
by considering or neglecting the gravity of the disc in
computations. Two cases are shown in each panel: First, the
gravitational field of the disc was computed in consistent manner with
the disc surface density $\sigmad(r)$ profile (solid line). Then, in
order to demonstrate the dependency on the disc mass more clearly, 
we increased the $\mu$ parameter
by factor of ten and recomputed the structure of
the cluster with this new value (dashed line). The corresponding change
of stellar density reaches order of several percent.
\label{fig:ratio}}
\end{figure}

The impact of disc gravity can be demonstrated by ``switching'' the 
gravitational field of the disc on and off in computations.
This is achieved by setting the value of $\mu$ parameter and
computing the corresponding $f_{\ast}(a;\mu)$. In 
Figure~\ref{fig:ratio}, we show resulting ratios
\begin{equation}
S(a;\mu)\equiv\frac{f_{\ast}(a;\mu)}{f_{\ast}(a;\mu=0)}
\label{eq:s_a_mu}
\end{equation} 
of two such distributions. Each curve was obtained with 
two identical models, except for the inclusion or omission 
of the disc gravity. The two frames differ only by the 
adopted model of the disc.
It turns out that perturbation to the stellar orbits by non-zero disc
mass speeds up the orbital decay and, consequently, it decreases the
number density of stars in the dragged cluster. This is visible even if the
mass ratio is small, i.e.\ of the order of $\mu\sim10^{-2}\div10^{-3}$.
Naturally, the influence of the disc gravity increases with $\mu$
increasing. Details of profile of the gravitational field
also matter: we see, for example, that the disc (iii) affects 
$f_{\ast}$ more than the disc (ii)+(iv) if their mass parameter
$\mu$ is comparable. The reason of this behaviour is that the former case
has a flatter curve of $\sigmad(r)$ (cp.\ Fig.~\ref{fig:kuzmin}) 
extending farther out from the black hole than the latter
case, and hence it exhibits stronger impact on the
cluster distribution.

Furthermore, we have included the orbital decay by gravitational
radiation. Indeed, it is highly desirable to estimate this type of
effects also because of forthcoming gravitational-wave experiments. We
employed a simple approximation (Peters \& Mathews 1963) and we found
that gravitational radiation influences the final $f_{\ast}(a)$ only
very near the centre (see \v{S}ubr \& Karas 1999 for details). For the
discs considered here, this occurs typically below $\la10\rg$ and,
therefore, it does not leave any remarkable imprint on the results.
Notice that this conclusion does not contradict the results of Narayan
(2000) who considered case of low-density flows, and therefore he found
the gravitational radiation to be more important for the orbital decay
than the hydrodynamical dissipation, also at larger distance from the
centre.

\begin{figure*} 
\includegraphics[width=\textwidth]{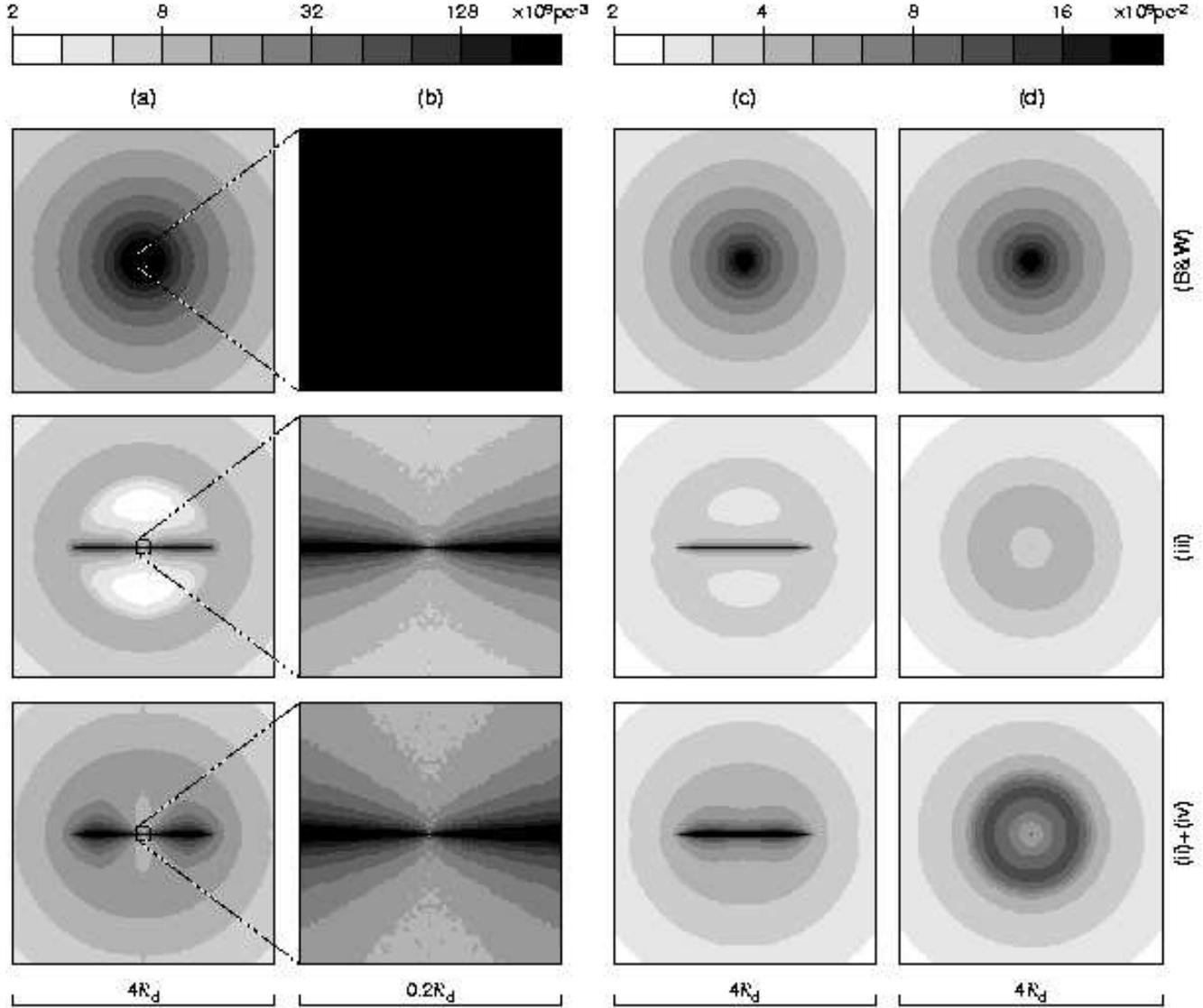}
\caption{The spatial density $n_\ast(r)$ and the corresponding projected
density of the cluster are shown using logarithmically spaced levels of
shading. Columns {\sf{}(a)} and {\sf{}(b)} represent the meridional
section at two different scales, namely, $4R_\idxs{d}$ and
$0.2R_\idxs{d}$ across (radii are expressed in terms of the disc outer
radius, $R_\idxs{d}$). Next columns are the edge-on {\sf{}(c)} and the
face-on {\sf{}(d)} projections of the cluster. Across columns, the upper
row shows the referential cluster ($n_{\ast}\,\propto\,r^{-7/4}$). In
subsequent rows, the system has been already modified via the
interaction with two types of discs, case (iii) and case (ii)+(iv), as
indicated.}
\label{fig:R_z_density}
\end{figure*}

Because the system is non-spherical but axially symmetric and
stationary, it is natural to portray its structure in a two-dimensional
spatial projection. Figure~\ref{fig:R_z_density} shows different
sections of the cluster arranged in four columns. Starting from the left
side, columns {\sf{}(a)} and {\sf{}(b)} represent the spatial number
density $n_{\ast}(r)$ in the meridional plane across the cluster.
Individual frames are centered on the black hole. Further, two columns
on the right show the projected (column) density, i.e.\ the total number
of stars integrated along the observer line of sight. Two different
lines of sight were chosen, perpendicular to {\sf{}(c)} and co-linear
with {\sf{}(d)} the symmetry axis. Interesting morphological patterns
can be revealed by comparing different views.

Let us describe the meaning of individual rows of
Fig.~\ref{fig:R_z_density}. In the first row, the density profile of the
unperturbed {\sf{}BW} distribution is plotted. It corresponds to the
outer cluster which is clearly spherical and serves as a starting
configuration for our computations.\footnote{Notice that $n_\ast(r)$
exceeds $1.7\times10^{11}\,\mbox{pc}^{-3}$ below $0.1R_\idxs{d}$. It is
thus out of density scale in the detailed plot of the central region --
column {\sf{}(b)}, top frame. However, such a high concentration is only
a formal artefact of the {\sf{}BW} model in which stellar density
diverges in the origin, $n_{\ast}(r)\,\propto\,r^{-7/4}$. The divergence
does not occur in our cluster interacting with the disc, in which case
the {\sf{}BW} solution is only used to describe the outer reservoir.} We
show this case for the sake of comparison with the structure of modified
clusters in subsequent rows. Again, we use the disc models (iii) and
(ii)+(iv) as two representative cases. 

Comparing different representations of the cluster one can clearly
observe the impact that star--disc collisions have on the cluster
structure. In particular, it is the increasing oblateness of the stellar
population in the core and, in some cases, the tendency to form an
annulus of stars. The effect is best seen in the bottom row in
Fig.~\ref{fig:R_z_density}. However, details of the modified cluster
structure obviously depend on the type of accretion disc that we choose. 
The reason of different structures stems from the different modes of
radial transport of the stars from the cluster to the central hole. In
the following paragraphs we examine several symptoms of the growing
deformation of the cluster.

\subsection{Anisotropy}
The rate of dragging the stars by the disc is higher for 
counter-rotating stellar orbits than for co-rotating or aligned ones,
hence, it is natural that the initial state of isotropy and sphericity
becomes gradually violated. Less obvious is the fact that toroidal
structures may be formed as a consequence. To see the effect we inspect
graphs of mean inclination $\avg{i}(a)$ (bottom row of
Fig.~\ref{fig:radial_profile}). The above-mentioned populations of
cluster members are clearly distinguished, in particular, small values
of $\avg{i}(a)$ indicate a flattened structure of the cluster with a
population of disc co-rotating orbits at small $a$. While the outer
cluster approaches an isotropic system ($\avg{i}\rightarrow\pi/2$) for
large $a$, the dragged cluster is clearly flattened to roughly a
constant value of mean inclination, ($\avg{i}\sim0.2)$. It drops
further in the embedded cluster where $\avg{i}\rightarrow0$. This
behaviour provides another substantiation for distinguishing different
populations that form the cluster. 

The limiting value of mean inclination can be derived by the following
simple analytical argument (which agrees with the results of
computations). Probability density of finding a body in
${\langle}a,a+{\rd}a\rangle\times{\langle}x,x+{\rd}x\rangle$ is
\begin{eqnarray}
 f_{\ast}(a,x) &{\bs}=&{\bs}\int \rd a_0\, \rd x_0\, \rd y_0\, \rd \omega_0\,
 \frac{f_0(a_0, x_0, y_0, \omega_0)}
 {t_{\rm r}(a_0)} \left| \frac{\rd a}{\rd t} \right|^{-1}  \nonumber \\
 &{\bs}\times& \delta\left( x - x(a; a_0, x_0, y_0, \omega_0) \right) ,
\end{eqnarray}
where integration is performed over the whole phase space of initial
values. The mean inclination is
\begin{equation}
\avg{i}(a) \equiv f_{\ast}^{-1}(a)\,\int\rd x\,\arccos{x}\,f_{\ast}(a,x)\,,
\label{eq:mean_inclination}
\end{equation}
where the normalization factor $f_{\ast}(a)$ is
\begin{equation}
f_{\ast}(a)=\int \rd x \, f_{\ast}(a,x).
\end{equation}

The mean inclination can be estimated analytically by neglecting gravity
of the  disc and assuming all orbits to be initially circular with
random inclination,
\begin{equation}
 f_0(a_0,x_0,y_0,\omega_0)=f_{\mbox{\tiny BW}}(a_0)\,\delta(y_0-1)\,.
\end{equation}
With the aid of eq.~(\ref{eq:circular_orbit}) the integral
in eq.~(\ref{eq:mean_inclination}) can be estimated in the form
\begin{equation}
\avg{i}(a) \sim \frac{2\sqrt{2}\,\pi}{2\ln{8} - 2\ln(1-x_{\rm max})-3 +
x_{\rm max}} \,.
\label{avgi}
\end{equation}
($x_{\rm max}$ denotes cosine of limiting inclination that distinguishes
stars of the dragged and embedded cluster). For thin disc models,
$1-x_{\rm max} \approx \frac{1}{2}(H/r)^2$ is roughly constant, reaching
values of the order of $10^{-7} \div 10^{-6}$. Hence, we obtain
\begin{equation}
\avg{i}(a) \approx -\frac{\sqrt{2}\,\pi}{\ln (1 - x_{\rm max})} \approx 0.3\,,
\end{equation}
which only slightly overestimates values we found by Monte Carlo
computations.

In the velocity space, anisotropy can be measured by means of factor
\begin{equation}
 \betaan\equiv1-
 \frac{{\langle}v_\idxt{T}^2{\rangle}}{2{\langle}v_\idxs{r}^2{\rangle}},
 \label{eq:beta_a}
\end{equation}
which is defined in terms of transverse and radial velocity components, 
$v_\idxt{T}$ and $v_\idxs{r}$, averaged over the cluster. Resulting
values of $|\betaan|$ are typically of the order of unity, zero value
corresponding to isotropic distribution. Averaged over the whole
cluster, our models give $\betaan\approx-1.5$, while values within
interval $(-5,-4)$ are obtained for the inner cluster.

\subsection{Diversification of eccentricities}
\begin{figure}
\includegraphics[width=\columnwidth]{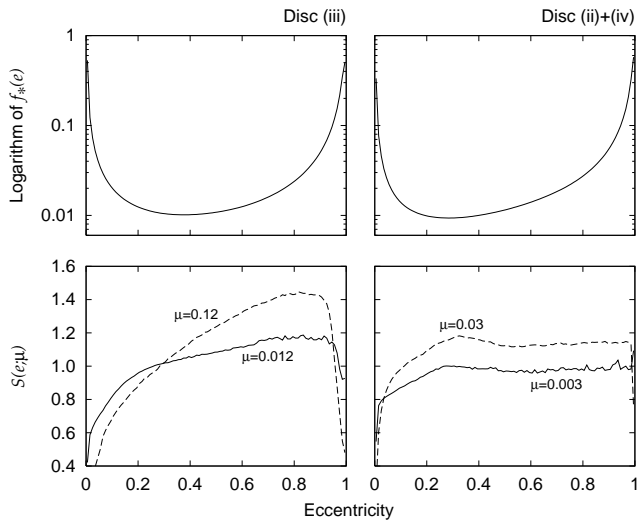}
\caption{Top: eccentricity distribution $f_\ast(e)$ in the inner 
cluster. Bottom: ratio $S(e;\mu)$ showing the effect of disc 
gravity according to eq.~(\ref{Deltan}). 
Parameters of the model are described in the text.}
\label{fig:eccentricity}
\end{figure}

By gradual circularization, every captured star contributes to the
resulting profile $f_{\ast}(e)\,{\rd}e$ proportionally to time it spends
within ${\langle}e,e+\rd{}e{\rangle}$. The orbital evolution slows-down
at the late phase and this leads to higher abundance of nearly circular
orbits. Prevailing circularisation is therefore anticipated because of
gradual energy losses and the resulting orbital decay. Fraction of
highly eccentric orbits in the cluster is small, yet it is important to
know because these orbits bring stars close to the black hole
where they can be preferentially captured or destroyed. The effect of
non-sphericity of the gravitational field complicates the long-term
tendency because it can lead to occasional increases of eccentricity.

We can distinguish two concurrent mechanisms that determine
eccentricities. Firstly, the selection rules
(\ref{eq:drag_vs_relax}) and (\ref{eq:geometry_condition}) define a loss
cone in the space of initial osculating elements. Stars on eccentric
orbits are continuously shipped to the inner cluster and this mechanism
accentuates the proportion of highly eccentric orbits with pericentres
below $R_\rd$, hence, such orbits are more abundant in the final
distribution. Secondly, it is the interaction of captured stars with the
disc which shapes the inner distribution. Combining the two mechanisms,
we obtain a double-peaked distribution with local maxima at low and at
high eccentricities and depression in the middle (see
Figure~\ref{fig:eccentricity}). This structure corresponds to an
intermediate phase of the cluster evolution which is seen in simulations
of Rauch (1995; cp.\ his Fig.~8). 

When the disc gravity is taken into account during the orbit
integration, resonances lead to eccentricity jumps, this way
contributing to highly eccentric orbits. We examined the overall impact
on the distribution function by means of ratio
\begin{equation}
S(e;\mu)\equiv\frac{f_{\ast}(e;\mu)}{f_{\ast}(e;\mu=0)}\;,
\label{Deltan}
\end{equation}
analogically to eq.~(\ref{eq:s_a_mu}). Figure~\ref{fig:eccentricity} 
suggests that a non-spherical perturbation to the gravitational field 
leads to higher abundance of more eccentric orbits.

\subsection{The mass stratification}
Let us now examine the mass spectrum of the cluster. We assume a
``canonical''
form of the initial mass function $\imf(M_\ast)$ (Salpeter 1955) 
together with the empirical relation for $\sigmas(M_{\ast})$ of
main sequence stars (e.g.\ Binney \& Merrifield 1998; Lang 1980):
\begin{equation}
\imf(M_\ast)\,\propto\,M_\ast^{-2.35},\quad
\sigmas(M_\ast)\,\propto\,M_{\ast}^{-0.6}.
\label{eq:salpeter}
\end{equation}
Stellar masses are generated in the range
$0.5M_{\sun}<M_{\ast}<10M_{\sun}$. Notice that the assumption
(\ref{eq:salpeter}) is needed for definiteness of the following example,
but its special form and the assumed range of masses are not required by
the adopted approach. This way we obtain profiles resembling, typically,
those for a cluster which is continuously supplied by equal-mass stars
at the lower boundary of the mass spectrum. The resulting distribution
follows partly from the fact that light stars represent a major fraction
of inserted bodies, partly it is due to their relatively slow migration
towards the centre. Also, radial transport of light bodies in the disc
is less efficient than that of heavy ones, except for very massive stars
which tend to open a gap in the disc. Once this happens, a qualitatively
different mode of radial motion takes place, typically orders of
magnitude slower than under the regime of density waves excitation. 

\begin{figure} 
\includegraphics[width=\columnwidth]{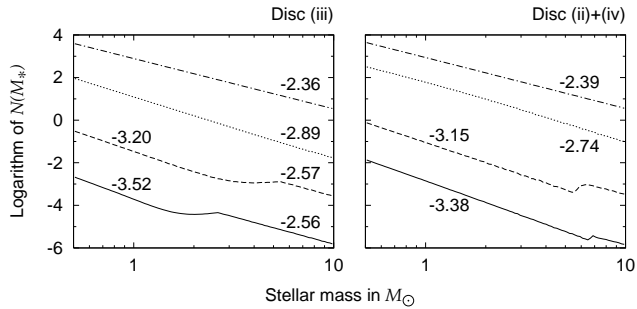}
\caption{An exemplary graph of the mass function
(with arbitrary normalization) for a cluster
modified by the interaction of its members with different
types of discs: case (iii) is shown on the left, case
(ii)+(iv) on the right. Individual curves represent the mass function 
of a sample of stars occurring below a given radius: 
$10^2\rg$ (solid), $10^3\rg$ (dashed), $10^4\rg$ (dotted)
and $10^5\rg$ (dash-dotted). Numbers are given with the 
curves to indicate the corresponding power-law indices of 
best-fit lines.}
\label{fig:mass_function}
\end{figure}

Gradual change of $\mmf(M_\ast)$ arises here due to unequal rate of
orbital decay of stars with different masses. The effect of mass
stratification can  be estimated by
\begin{equation}
{\mmf}(M_\ast)\propto\imf(M_\ast)\;|\dot{a}(M_\ast)|^{-1}.
\label{eq:mmf}
\end{equation}
See Figure~\ref{fig:mass_function} for profiles resulting from
computations (values of the best-fit power-law index are given in the
plot). Different curves show the stellar-mass distribution within the
corresponding radial distance from the centre.

Let us discuss case of the disc (ii)+(iv) in 
Fig.~\ref{fig:mass_function} in more detail. Majority of stars in the
region below $10^5\rg$ belong to the outer cluster. The mass function
remains almost unmodified in this region. However, selecting only stars
inside $10^4\rg$ into consideration, substantial fraction of orbits are
already aligned with the disc. Their radial velocity of migration is
proportional to $\Sigma_\ast^{-1}$ and, through this dependence, it is
$\propto{}M_\ast^{0.6}$. Hence, employing eq.~(\ref{eq:mmf}) we estimate
that the mass function index decreases by $0.6$ and it reaches the value
$-2.95$.  Going still closer to the centre, we find more stars that are
fully embedded in the disc and migrate in the regime of density waves,
i.e.\ with radial velocity $\propto\,M_\ast$. They cause a further drop
of the mass index and the expected value in this case is $-3.35$. This
trend is in good agreement with the behaviour we see in numerical
simulations.

A modification of the power-law profile occurs at the upper boundary of
the stellar mass range as a consequence of changing the mode of radial
migration. This is evident in the case (iii). Considering
$r{\la}10^3\rg$, majority of stars again belong to the embedded cluster,
but massive stars now succeed to open a gap in the disc and, therefore,
they continue migrating to the centre with the radial velocities
independent of $M_{\ast}$. For this reason the mass-function index is
raised closer to its initial value. Within the innermost region (below
$100\rg$) the limiting mass for the gap opening decreases and the 
transition between two different slopes moves to lower $M_\ast$.

\begin{figure} 
\includegraphics[width=\columnwidth]{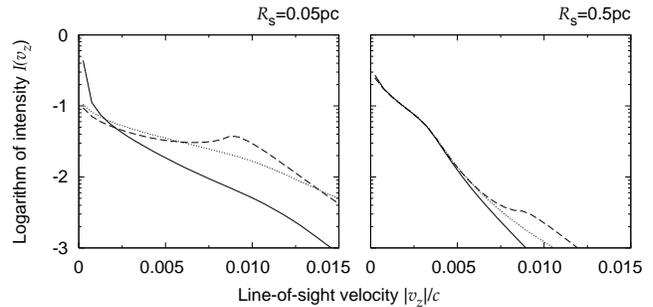}
\caption{Velocity profile along the line of sight of the inner cluster,
integrated across a column of cross-sectional radius
$R_\idxs{s}=10^4\rg$ (left panel) and $R_\idxs{s}=10^5\rg$ (right
panel). Solid and dashed lines represent different inclinations of the
observer: $\xi=0\degr$ and $60\degr$, respectively. Growing anisotropy 
of the modified cluster produces the dependence of measured line profile
on observer's view angle, i.e.\ $I{\equiv}I(v_z;\xi)$. The thin dotted
line stands for the referential {\sf{}BW} distribution, which is
spherically symmetric.}
\label{fig:line_detail_M2}
\end{figure}

\subsection{Velocity dispersion}
Let us examine integral properties of the cluster that reflect the state
of the inner region in terms of potentially observable quantities: the
shape of a synthetic spectral line $I(v_z)$ (i.e.\ intensity as a
function of line-of-sight velocity $v_z$ near the projected center of
the cluster) and velocity dispersion $\sigma_z$ along line of sight.
It is worth noting that similar spectral profiles were computed for the
reference cluster already in the original Bahcall \& Wolf (1976) paper,
so one can check how the picture is changed by the presence of the
assumed disc and what imprints could be traced in the integrated lines
from the nucleus unresolved to individual stars. 

In Figure~\ref{fig:line_detail_M2}, the left panel shows the
line-of-sight profile that was obtained by averaging over a column of
radius $R_{\rm{}s}=10^4\rg$\label{p:r_s}, centered on the cluster core.
This profile determines an infinitesimally narrow line (in a local 
frame of each star). We neglected all other effects that might influence 
the final spectral feature and assumed the disc (iii) with $R_\idxs{d}=10^4\rg$.
Another view of the same system is shown in the right panel, where we
chose lower spatial resolution for comparison. Also in these
profiles one can recognize the tendency of the stellar sub-system to be
flattened. Stellar vertical velocity component decreases with respect to
the component parallel with the equatorial plane, which manifests itself
by gradual decrease of the line width as the view angle becomes parallel
with the symmetry axis. Local maximum of the line occurs around
$v\sim\sin(\xi)v_\idxt{K}(R_\idxs{d})$. In some cases, this secondary
peak may exceed the central maximum and dominate the whole profile.
High-velocity tails of the line profiles are also noticeably affected.

\begin{figure} 
\includegraphics[width=\columnwidth]{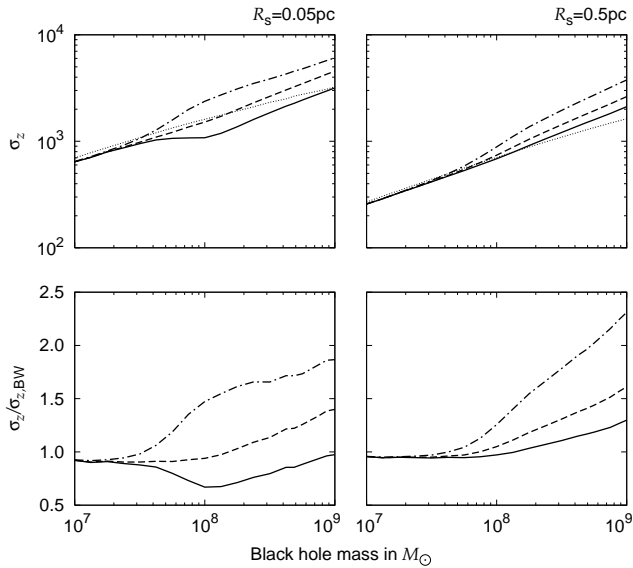}
\caption{Velocity dispersion $\sigma_z$ along the line of sight as a function 
of the central mass $\mbh$ (top panels). Different line-types represent 
observer's view angle: $\xi = 0\degr$ (solid), $30\degr$ (dashed) and 
$90\degr$ (dash-dotted). For the sake of comparison, thin dotted line
is an unperturbed profile of the outer cluster. In bottom panels
we magnify the differences showing the ratio of $\sigma_z$
with respect to the {\sf{}BW} case.}
\label{fig:disperze_M2}
\end{figure}

In Figure~\ref{fig:disperze_M2} we show predicted dispersion $\sigma_z$
using a circular aperture of radius $R_\idxs{s}$. In the case of initial
{\sf{}BW} distribution, $\sigma_z$ scales roughly proportionally to
$\mbh^{1/2}$. The slope is somewhat modified when the range of
semi-major axes is restricted in the reference cluster (in particular,
we assumed $R_\idxs{d}{\la}a_0{\la}\rh$). If the cluster is modified by
interaction with the disc, a threshold mass can be found in the
$\sigma_z(\mbh)$ dependence. Its value depends on the fraction of stars
belonging to the inner cluster. At the lower end of the $\mbh$ range,
dispersion resembles the previous unperturbed case, because the
anisotropic inner cluster is entirely concealed by the outer one. On the
other side, for large $\mbh$ we observe a steeper slope of the
$\sigma_z(\mbh)$ relation -- a consequence of the newly established
radial density profile $n_{\ast}(r)$. The exact form of this relation
depends on the view angle, which is another manifestation of the inner
cluster anisotropy.

\subsection{Mass inflow from the outer cluster}
\label{mass_flow}
The rate of mass inflow in stars, $\Mdot$, characterizes their radial
drift from the outer reservoir. Related to $\Mdot$ is the induced energy
dissipation ${\Kdot}$, also caused by star--disc collisions. These two
quantities can be expressed in a common way. To this aim we introduce 
a new variable, $\cal A$, denoting a quantity that is formed by contributions 
carried by individual stars, as they sink to the centre ($\Kdot$ or $\Mdot$
are just two examples of such quantity). The corresponding flow
rate is
\begin{eqnarray}
 \dot{\cal A} &{\bs}=& {\bs}\!\int_{a_{\rm{}min}}^{a_{\rm{}max}}
 \!\!\!\!\!\!{\rd}a_0 
 \int_0^1\!\!\!{\rd}e_0 
 \int_{-1}^1\!\!\!{\rd}x_0
 \int_{0}^{\pi}\!\!\!{\rd}\omega_0 \;
 {\cal A}\;f_{\mbox{\tiny BW}}(a_0, e_0)\; t_{\rm r}^{-1}(a_0)
 \nonumber \\
 &{\bs}\times& {\bs\!} 
 \Theta\!\left(-\dot{a}_{0} - a_0/t_{\rm r} \right) \;
 \Theta\!\left(R_\rd - R_0 \right)\!,
\label{eq:dot_integral} 
\end{eqnarray}
where the two theta-functions limit the region of parameter space from
which the stars are captured by the disc; cf.\ conditions
(\ref{eq:drag_vs_relax}) and (\ref{eq:geometry_condition}). We obtain
$\Mdot$ by setting
\begin{equation}
{\cal A}\rightarrow M_\ast
\end{equation}
in the integrand of eq.~(\ref{eq:dot_integral}). Likewise we obtain
${\Kdot}$ for
\begin{equation}
{\cal A}\rightarrow \Delta E \equiv G\,\mbh\, M_\ast
\left(\frac{1}{a_\idxs{f}} - \frac{1}{a_0}\right).
\end{equation}
In both cases we performed integration numerically. 
One can omit the leading theta-function  in
eq.~(\ref{eq:dot_integral}) because the drag time is shorter than
relaxation time for majority of initial conditions (we assumed customary
values of accretion disc parameters, $\dot{M}_{0.1}=1$,
$\alpha_{0.1}=1$). Even if relaxation cannot be immediately neglected,
we may still ignore the mentioned term in eq.~(\ref{eq:dot_integral})
and substitute $R_\idxs{d}{\rightarrow}R_\idxs{i}$.

Further, we write analytical estimates for $\Mdot$ and ${\Kdot}$,
\begin{equation}
 \Mdot = 1.7 \times 10^{-2} M_8^{5/4} \, n_6^2 \, 
 \left( \frac{M_\ast}{M_\odot} \right)^2
 \!\!\frac{R_\rd}{10^4 \rg}\; {\cal F} \; \left[\frac{M_\odot}{\rm yr}\right]
\label{eq:M_dot_estimate}
\end{equation}
\begin{equation}
 {\Kdot} = 10^{-4} M_8^{5/4} \, n_6^2 \, 
 \left( \frac{M_\ast}{M_\odot} \right)^3
 {\cal E} \quad \left[\frac{M_\odot c^2}{\rm yr} \right] ,
 \label{eq:E_dot_estimate}
\end{equation}
where $\cal E$ and $\cal F$ are order-of-unity and slowly varying
functions for
$0.01R_\idxs{d}{\la}a_{\rm{}min}{\la}R_\idxs{d}$,
$R_\idxs{d}{\la}a_{\rm{}max}{\la}100R_\idxs{d}$.

The total accretion rate onto the black hole is a sum of $\Mdot$, which
involves stars from the cluster, and the accretion rate $\dot{M}$ of the
gas in the disc. As for ${\Kdot}$, we expressed this quantity as
difference $V_{\rm{}i}-V_{\rm{}f}$ of the initial and final potential
energy of a star, where $V_{\rm{}f}$ was estimated from
eq.~(\ref{eq:circular_orbit}). 

We remind the reader that a uniform distribution of initial 
eccentricities was assumed, but this limitation is not very important
because only a narrow range of large initial
eccentricities populate the resulting distribution.\footnote{For 
another form of initial
eccentricity distribution, an estimate of the flow of mass and kinetic
energy  can be obtained by multiplying formulae
(\ref{eq:M_dot_estimate}) and (\ref{eq:E_dot_estimate}) by factor
$f_{\ast}(e_0=1)$. We verified accuracy of this scaling in two cases,
$f_{\ast}(e_0)=2e_0$ and $f_{\ast}(e_0)=0.58\exp(e_0)$, and we found
that it indeed agrees with numerical computations.} Eqs.\
(\ref{eq:M_dot_estimate}) and (\ref{eq:E_dot_estimate}) can therefore 
serve as useful order-of-magnitude estimates. 

\section{Discussion and conclusions}
\label{conclusions}
We have developed the model of a cluster departing from its initial 
sphericity and isotropy as a result of star--disc collisions. One can
anticipate various observational consequences of this interaction, and
several of them were examined in Sec.~\ref{results}. Hereafter we list
other implications but we defer detailed comparisons with actual
evidence until a more complete picture is available. Let us remark that
observations can be confronted in two areas: firstly, it is the overall
form of the stellar distribution in the nuclear cluster as discussed
above (which however corresponds to sub-parsec scales, not yet
sufficiently well resolved), and then it is a discussion of individual
events of star--disc collisions or disruptions, whose frequency is
linked with the rate of orbital decay and which may be detected by a
sudden releases of X-rays or by gravitational-waves emission.

The orbital decay of stars near a black hole is indeed
relevant for forthcoming gravitational wave experiments, because the
gas-dynamical drag needs to be taken into account in order to predict
waveforms with sufficient accuracy. It has been estimated (Narayan 2000;
Glampedakis \& Kennefick 2002) that this effect can be safely ignored at
late stages, shortly before the star plunges into the hole, {\it if\/}
accretion takes place in the mode of a very diluted flow. 
However, the situation is quite different in case of AGN hosting
rather dense nuclear discs (\v{S}ubr \& Karas 1999). Consistent
evaluation of gravitational radiation from dense stellar systems would
require reformulation of our model within the general relativity
framework, which is beyond the scope of this paper, but we estimate that
the motion of only a minor fraction of stars very near the centre should
be noticeably affected by the emission of gravitational waves. One can
expect that exact calculation would lead to even a weaker effect because
our stellar system tends to a flattened rotating configuration that does
not radiate in the limit of perfect axial symmetry 
(luminosity emitted in gravitational radiation by particles orbiting in a
ring decreases exponentially with the number of particles; Nakamura \&
Oohara 1983). In other words, gas-dynamical effects most likely dominate
over the gravitational radiation as far as their influence on the cluster 
structure is concerned.

The modified cluster structure is obviously pertinent for various
studies of the black-hole feeding problem and, vice-versa, for the issue
of a possible feedback that a super-massive black hole may exhibit on
the the host galaxy bulge. However, the well-known empirical relations
between the black hole mass and the bulge properties apply to the
region substantially exceeding $\rh$. Typically,
${\mbh}/M_{\rm{}bulge}\sim10^{-3}$, while the two masses should be
comparable where the cluster structure is to be directly influenced by
the central hole, which is what we assumed here. An interesting
direction in which our model can be further developed therefore
concerns a more realistic form of the outer cluster and the way it is
connected with the galaxy bulge. This could help to build a model for
the $\sigma_{\rm c}(\mbh)$ relation extending beyond the domain
of black hole gravitational influence. In this context it is worth to
mention a recent idea (Miralda-Escud\'e \& Kollmeier 2004) that
star--disc collisions can indeed be seen as a self-regulating process
that helps to feed the central black hole and control its growth.

Inherent limitations persist in our present discussion, namely, we have
not incorporated a fully self-consistent treatment of the disc gravity.
Even though we computed the gravitational field across a sufficiently
large domain of space, we did not account for the feedback which
star--disc collisions exert on the disc structure. Introducing some kind 
of a clumpy model of the disc will be very interesting, as it may exert a
more substantial effect on the cluster structure by elevating the impact
of star--disc collisions. Different configurations of a clumpy disc have
been proposed (Kumar 1999; Fukuda, Habe \& Wada 2000; Hartnoll \&
Blackman 2001; Kitabatake \& Fukue 2003) which will be relevant for this
investigation. 

Also relevant is the idea of enhanced star
formation caused by supernovae that are triggered in a self-gravitating
disc, changing substantially its structure
(Collin \& Zahn 1999). We can expect that the inner cluster will be
affected in different way in those galactic nuclei where nuclear
starbursts occur simultaneously with AGN phenomenon (Tenorio-Tagle et
al.\ 2003; Gonz\'ales Delgado et al.\ 2004). The circumnuclear starburst
phenomenon has been indeed indicated in some AGN on $\sim10^2\;$pc
scales, which could correspond to the outskirts of a dusty torus surrounding
the central black hole (e.g.\ Aretxaga et al.\ 2001; Heckmann et al.\
1997; Schinnerer et al.\ 2000). Star formation could then take place at
outer parts of the nuclear cluster.

\section*{Acknowledgements}
We are grateful to M.~Freitag for helpful discussions on the importance 
of gravitational relaxation in the nuclear cluster. We thank A.~Kawka
and W.~Klu\'{z}niak for reading the manuscript and for comments. An 
anonymous referee provided us with a number of very
useful suggestions improving the paper. We acknowledge
financial support from the postdoctoral grant 205/02/P089 and the
research project 205/03/0902 of the Czech Science Foundation, as well as
the project 299/2004 of the Charles University in Prague.

\label{lastpage}
\end{document}